\documentclass[twoside]{article}
\usepackage[latin9]{inputenc}
\usepackage{amsmath}
\usepackage{amssymb}

\makeatletter

\let\epsilon=\varepsilon

\newtheorem{defi}{Definition}[section]

\newtheorem{theorem}{Theorem}[section]
\newtheorem{lemma}{Lemma}[section]

\def\tire{\thinspace--\thinspace}
\date{}

\title{Harmonic Chain with Weak Dissipation}

\author{A.A.~Lykov\and V.A.~Malyshev 
\thanks{Moscow State University, Faculty of Mechanics and Mathematics, 
Vorobievy Gory, 119991 Moscow}}

\makeatother

\begin{document}
\maketitle
\begin{abstract}
We consider finite harmonic chain (consisting of $N$ classical particles)
plus dissipative force acting on one particle (called dissipating
particle) only. We want to prove that ``in the generic case'' the
energy (per particle) for the whole system tends to zero in the large
time limit $t\to\infty$ and then in the large $N$ limit. ``In the
generic case'' means: for almost all initial conditions and for almost
any choice of the dissipating particle, in the thermodynamic limit
$N\to\infty$. 
\end{abstract}

\section{Introduction}

The energy $H(t)$ in Hamiltonian systems is conserved. After adding
dissipation terms, $H(t)$ is not conserved anymore but becomes a
non-increasing function. We consider finite harmonic chain (consisting
of $N<\infty$ classical particles) plus dissipative force acting
on one particle (having number $n$ and called dissipating particle)
only. It appears that $H(N,n,t)$ is very irregular with respect to
$N$ and $n$. However, we want to prove that in the general case
the energy per particle $N^{-1}H(N,n,t)$ tends to zero in the limit
where first $t\to\infty$ and then $N\to\infty$. ``In the general
case'' means: for almost all initial conditions and for almost any
choice of the dissipating particle, see exact definitions below.

Finite linear hamiltonian systems (including also additional linear
terms) were studied by many authors, see for example \cite{Kozlov,Williamson,Krejn_1,Krejn_2},
but with the goals different from ours. The papers \cite{Chern,Bilir}
are closer to ours, in \cite{Bilir} the authors study the cases when
the dimension of $L_{0}$ (see next section) is zero.

Our main formulas concern linear algebra and our main conventions
about notation are as follows: all vectors are row vectors, there
will be introduced two scalar products $(,)_{k},k=1,2,$ and we denote
the linear span of the set $R$ of vectors by $\langle R\rangle$.

\section{Main results}

We consider the system of $N$ particles with the phase space 
\[
L=L_{2N}=\mathbb{R}^{2N}=\{\psi=(q,p):\ q=(q_{1},\ldots,q_{N}),p=(p_{1},\ldots,p_{N})\in\mathbb{R}^{N}\},
\]
with the scalar product 
\[
(\psi,\psi')_{1}=\sum_{i=1}^{N}(q_{i}q_{i}'+p_{i}p_{i}')
\]
and with quadratic Hamiltonian (energy) 
\[
H(\psi)=T+U,\quad T=\frac{1}{2}\sum_{i=1}^{N}p_{i}^{2},\quad U=\frac{1}{2}\sum_{i,j}V(i,j)q_{i}q_{j},
\]
where $V=(V(i,j))$ is a (symmetric) positively definite $(N\times N)$-matrix.
We assume unit masses of the particles. The phase space is the direct
sum of the orthogonal coordinate and momentum subspaces, $l_{N}^{(q)}$
and $l_{N}^{(p)}$ correspondingly, with the induced scalar products
$(q,q')_{1}$ and $(p,p')_{1}$.

The dynamics of this system is defined by the following system of
equations, $i=1,\ldots,N$: 
\begin{align*}
\frac{dq_{i}}{dt} & =p_{i},\\
\frac{dq_{i}}{dt} & =-\frac{\partial U}{\partial q_{i}}-\alpha p_{i}\delta_{in}=-\sum_{j}V(i,j)q_{j}-\alpha p_{i}\delta_{in},
\end{align*}
where we appended a dissipation force for one particle ONLY, that
is, we fix particle with number $n$ and assume that it is subjected
to dissipation with fixed $\alpha>0$. More generally, such system
can be written in the matrix form 
\[
\dot{\psi}=A\psi,\quad\psi\in L,
\]
with $(2N\times2N)$-matrix 
\[
A=\left(\begin{matrix}0 & E\\
-V & -D
\end{matrix}\right)
\]
with the matrices $V,D$ and the unit matrix $E$ acting in $\mathbb{R}^{N}$.
The dissipation matrix $D\geqslant0$ is symmetric and non-negative,
in our case it is diagonal with the only non-zero element. The dynamics
can also be presented as the equivalent linear second order ODE system
in $\mathbb{R}^{N}$: 
\[
\ddot{q}+D\dot{q}+Vq=0,\quad q=(q_{1},\ldots,q_{N})\in\mathbb{R}^{N}.
\]
For any initial $\psi(0)\in L$ the solution is $\psi(t)=e^{tA}\psi(0)$.
It is well-known (and easy to check) that the energy is non-increasing
and 
\begin{equation}
\frac{d}{dt}H(\psi(t))=-(Dp,p)_{1}.\label{non-increasing}
\end{equation}

\begin{defi} 
$L_{0}\subset L$ is the subset of elements $\psi$ of the phase space
$L$ for which the energy is conserved, 
\[
L_{0}=\Bigl\{\psi\in L:\ \frac{d}{dt}H(e^{tA}\psi)=0,\ \forall t>0\Bigr\}.
\]
$L_{-}\subset L$ is the subset of elements $\psi$ of the phase space
$L$ for which the energy tends to zero as $t\to\infty$, 
\[
L_{-}=\{\psi\in L:\ H(e^{tA}\psi)\to0,\ t\to\infty\}.
\]
\end{defi}

\begin{theorem} \label{t1} Subsets $L_{0}$ and $L_{-}$are orthogonal
linear subspaces of $L$, invariant with respect to the dynamics,
and $L$ is their direct sum, 
\[
L=L_{0}\oplus L_{-}.
\]
If $\psi\in L_{-}$, then $H(e^{tA}\psi)\to0$ exponentially fast.
More exactly, for some constants $c_{1},c_{2}>0$ ($c_{1}=c_{1}(\psi)$
depends on $\psi$), 
\[
H(e^{tA}\psi)\leq c_{1}\exp(-c_{2}t)
\]
\end{theorem}

Our main result concerns the chain of harmonic oscillators with the
Hamiltonian 
\[
H=\frac{1}{2}\sum_{k=1}^{N}p_{k}^{2}+\frac{\omega_{0}}{2}\sum_{k=1}^{N}(x_{k}-ka)^{2}+\frac{\omega_{1}}{2}\sum_{k=2}^{N}(x_{k}-x_{k-1}-a)^{2},
\]
where $\omega_{0},\omega_{1},a>0$. In terms of the deviations 
\[
q_{k}=x_{k}-ka
\]
the Hamiltonian becomes 
\[
H=\frac{1}{2}\sum_{k=1}^{N}p_{k}^{2}+\frac{\omega_{0}}{2}\sum_{k=1}^{N}q_{k}^{2}+\frac{\omega_{1}}{2}\sum_{k=2}^{N}(q_{k}-q_{k-1})^{2}.
\]
We shall denote by $\gcd(a_{1},a_{2})$ the greatest common divisor
of the natural numbers $a_{1},a_{2}$.

\begin{theorem} \label{t2} The dimension $D_{n}(N)=\mathrm{dim}L_{0,n}$
enjoys the properties (\ref{divisor})--(\ref{averaging_N}). 
\begin{equation}
D_{n}(N)=\gcd(N,2n-1)-1.\label{divisor}
\end{equation}

For any $\epsilon>0$ there exists some $c(\epsilon)$ such that 
\begin{equation}
S(N)=\frac{1}{N}\sum_{n=1}^{N}D_{n}(N)\leq c(\epsilon)N^{\epsilon}.\label{averaging_n}
\end{equation}
as $N\to\infty$. 

For some absolute constant $c>0$ and any $N_{0}>0$, 
\begin{equation}
\frac{1}{N_{0}}\sum_{N\leqslant N_{0}}\frac{S(N)}{N}<c\ln N_{0}.\label{averaging_N}
\end{equation}
\end{theorem}

The following examples show that the function $D_{n}(N)$ is very
irregular in $n$ and $N$: 
\begin{itemize}
\item If $N=2^{k}$ for some $k>0$, then $D_{n}(N)=0$ for all $n$. 
\item If $N=2m-1$ and $n=m$, then $D_{n}(N)=2m-2=N-1$. 
\end{itemize}
Note that $S(N)$ is the mean dimension if the particle $n$ is chosen
randomly, and $N^{-1}S(N)$ is the mean energy per particle. Theorem~\ref{t2}
shows that if the initial energies of the particles in the system
are uniformly bounded as $N\to\infty$, and thus the initial energy
of the whole system is not more than of order $N$ (that is, proportional
to the number of particles), then the energy remaining in the system
after a long time is very small, namely of the order $N^{\epsilon}$,
and the energy per particle is of order $N^{-1+\epsilon}$. It can
be said, roughly speaking, that in the thermodynamic limit $N\to\infty$
the remaining mean energy per particle is zero (even if the kinetic
energy is pumping out only of a single particle).

\section{Proofs}

\subsection{Proof of Theorem~\ref{t1}}

Define the second scalar product in $L$ by 
\[
(\psi,\psi')_{2}=(Vq,q')_{1}+(p,p')_{1},
\]
and denote by $e_{n}=(0,\ldots,0,1,0,\ldots,0)\in l_{N}^{(p)}$ the
coordinate vector with $1$ at the $n$th place. Define also the vector
$g_{n}=(0,e_{n})\in L$.

We shall prove that 
\begin{equation}
L_{0}=\{\psi\in L:\ (e^{tA}\psi,g_{n})_{1}=0,\ \forall t>0\}=\{\psi\in L:\ (e^{tA}\psi,g_{n})_{2}=0,\ \forall t>0\}.\label{L_0_representations}
\end{equation}
Indeed, we have 
\[
\frac{d}{dt}H(e^{tA}\psi)=-\alpha p_{n}^{2}(t)=-\alpha(e^{tA}\psi,g_{n})_{1}^{2}.
\]
This is just rewriting of the right-hand side of (\ref{non-increasing}),
\[
(Dp,p)_{1}=\alpha p_{n}^{2}(t),\qquad p_{n}(t)=(e^{tA}\psi,g_{n})_{1}.
\]

Linearity of $L_{0}$ follows from (\ref{L_0_representations}), the
invariance of $L_{0}$ follows by differentiation of the equality
$(e^{tA}\psi,g_{n})_{1}=0$.

Define a subspace $l_{V}\subset\mathbb{R}^{N}$ as the linear span
of the vectors $V^{j}e_{n},\ j=0,1,\ldots$, and introduce the subspaces
$L^{(k)}$ as the orthogonal complements to $L_{0}$ with respect
to the scalar products $(,)_{k},\ k=1,2$ correspondingly.

\begin{lemma} \label{l1} $\phantom{aaa}$ 
\begin{itemize}
\item {\rm 1.} $L^{(1)}=L^{(2)}$. 
\item {\rm 2.} $L^{(2)}=\{(q,p)\in L:\ q,p\in l_{V}\}$. 
\item {\rm 3.} $L^{(2)}$ is invariant with respect to $A$. 
\end{itemize}
\end{lemma} \textbf{Proof.}\ We will use the following identity
\[
A=IQ-\alpha\Gamma,
\]
where $I,Q,\Gamma$ are matrices in $R^{2N}$ defined by 
\[
I((q,p))=(p,-q),\quad Q((q,p))=(Vq,p),\quad\Gamma\psi=(\psi,g_{n})_{1}g_{n}.
\]
Note that 
\[
(\psi,\psi')_{2}=(Q\psi,\psi')_{1}.
\]
Let $A_{k}^{*}$ be the adjoint operator to $A$ with respect to the
corresponding scalar products $(,)_{k},\ k=1,2$. It is easy to see
that 
\begin{align*}
A_{1}^{*}= & \ -QI-\alpha\Gamma,\\
A_{2}^{*}= & \ Q^{-1}A_{1}^{*}Q=-IQ-\alpha\Gamma.
\end{align*}
For $k=1,2$ put 
\[
M_{k}=\langle\{(A_{k}^{*})^{j}g_{n}\}_{j=0,1,\ldots}\rangle.
\]
We will prove that the orthogonal complements (with respect to the
corresponding scalar products) $M_{k}^{\perp}=L_{0},\ k=1,2$. Let
$\psi\in M_{k}^{\perp}$. Since 
\begin{equation}
(e^{tA}\psi,g_{n})_{k}=(\psi,e^{tA_{k}^{*}}g_{n})_{k}=\sum_{j=0}^{+\infty}\frac{t^{j}}{j!}\,(\psi,(A_{k}^{*})^{j}g_{n})_{k}=0\label{series}
\end{equation}
for any $t>0$, we have $\psi\in L_{0}$. Vice-versa, let $\psi\in L_{0}$.
Then for any $j=0,1,\ldots$ we have a chain of equalities 
\[
0=\frac{d^{j}}{dt^{j}}\,(e^{tA}\psi,g_{n})_{k}\vert_{t=0}=(A^{j}\psi,g_{n})_{k}=(\psi,(A_{k}^{*})^{j}g_{n})_{k}.
\]
It follows that $\psi\in M_{k}^{\perp}$, and thus $M_{k}^{\perp}=L_{0}$,
which is equivalent to 
\[
M_{1}=L^{(1)},\qquad M_{2}=L^{(2)}.
\]
Let us use induction in $m$ to prove that 
\[
\langle\{(A_{2}^{*})^{j}g_{n}\}_{j=0,1,\ldots,m}\rangle=\langle\{(IQ)^{j}g_{n}\}_{j=0,1,\ldots,m}\rangle.
\]
For $m=0$ the statement is evident. The inductive hypothesis is:
for some $c_{j}$, 
\[
(A_{2}^{*})^{m}g_{n}=\sum_{j=0}^{m}c_{j}(IQ)^{j}g_{n}.
\]
Applying to both sides of this equality the operator $A_{2}^{*}=-IQ-\alpha\Gamma$,
we get 
\[
(A_{2}^{*})^{m+1}g_{n}=\!-\sum_{j=0}^{m}c_{j}(IQ)^{j+1}g_{n}-\alpha((A_{2}^{*})^{m}g_{n},g_{n})_{2}g_{n}\in\!\langle\{(IQ)^{j}g_{n}\}_{j=0,1,\ldots,m+1}\rangle.
\]
Similarly, one can prove the inverse inclusion for the corresponding
linear spans. One can prove similarly that 
\[
L^{(1)}=\langle\{(QI)^{j}g_{n}\}_{j=0,1,\ldots}\rangle.
\]
It is easy to check that the following identities hold for $j=0,1,\ldots$
\begin{align*}
(IQ)^{2j}g_{n}= & \ (-1)^{j}(0,V^{j}e_{n}),\\
(IQ)^{2j+1}g_{n}= & \ (-1)^{j}(V^{j}e_{n},0).
\end{align*}
Thus, $L^{(2)}=l_{V}\oplus l_{V}$. Similarly for $j=0,1,\ldots$
one can check the formulas 
\[
(QI)^{2j}g_{n}=\ (-1)^{V}(0,V^{j}e_{n}),
\]
\[
(QI)^{2j+1}g_{n}=\ (-1)^{j}(V^{j+1}e_{n},0).
\]
It follows that $L^{(1)}=V(l_{V})\oplus l_{V}$, where $V(l_{V})$
is the image of $l_{V}$ under the mapping $V$. Since $l_{V}$ is
invariant with respect to $V$ and $V$ is invertible, $V(l_{V})=l_{V}$.
Thus, $L^{(1)}=l_{V}\oplus l_{V}=L^{(2)}$.

Invariance with respect to the operator $A$ follows from relations
\begin{align*}
A(V^{j}e_{n},0)^{T}= & \ (0,-V^{j+1}e_{n})\in L^{(2)},\\
A(0,V^{j}e_{n})^{T}= & \ (V^{j}e_{n},0)-\alpha(V^{j}e_{n},e_{n})_{1}g_{n}\in L^{(2)}.
\end{align*}
The lemma is proved. \ \ \ $\square$

Now we will use the classical result about ordinary differential equations.
Namely, consider the following system 
\begin{equation}
\dot{y}=f(y),\quad y\in\mathbb{R}^{n}.\label{*}
\end{equation}
If a function $v(y)\in C^{1}$ is given, we will denote its derivatives
along the trajectories as 
\[
\frac{dv}{dt}=\sum_{k=1}^{n}\frac{\partial v}{\partial y_{k}}(y)f_{k}(y).
\]
\begin{theorem}[Barbashin\tire Krasovskij's theorem
\cite{Barbashin}, p.\,19, Thm.\;3.2] 
\label{t3} Let $f$ be continuous, $f(0)=0$ and for $|y|\leqslant\rho$
$(\rho>0)$ there exists a (Lyapounov) function $v(y)\in C^{1}$ such
that $v(0)=0,$ $v(y)>0$ for $y\ne0$, $dv/dt\leqslant0$. Assume
moreover that the set of all $y$ for which $dv/dt=0$ does not contain
any whole trajectory, except $y=0$. Then the zero solution of (\ref{*})
is asymptotically stable. \end{theorem}

\begin{lemma} \label{l2} The following equality holds 
\[
L_{-}=L^{(2)}.
\]
\end{lemma} Let us prove that $L^{(2)}\subset L_{-}$. We use Lemma~\ref{l1}
and choose $v(\psi)=H(\psi)$. Then 
\[
\frac{dv}{dt}(\psi)=-(Dp,p)_{1}=-\alpha(\psi,g_{n})_{2}^{2}\,.
\]
The trajectories which belong to the set of zeros of $dv/dt$, by
definition belong to $L_{0}$. As $L_{0}\cap L^{(2)}=\{0\}$, the
zero trajectory is asymptotically stable on $L^{(2)}$. This means
that for any $\psi\in L^{(2)}$ we have $H(e^{tA}\psi)\to0$ as $t\to+\infty$.
Let us prove the inverse inclusion. Note that for any $\psi\in L$
we have $H(\psi)=(\psi,\psi)_{2}$. Let $\psi\in L_{-}$. We have
\[
\psi=\psi^{(0)}+\psi^{(2)},\quad\psi^{(0)}\in L_{0},\ \psi^{(2)}\in L^{(2)}.
\]
Then $H(e^{tA}\psi)=H(\psi^{0})+H(\psi^{(2)})\to H(\psi^{0})=0$ by
definition of $L_{-}$. It follows that $\psi^{(0)}=0$. Lemma~\ref{l2}
and thus Theorem~\ref{t1} are proved.

\subsection{Proof of Theorem~\ref{t2}}

First, we need the following general assertion. \begin{lemma} \label{l3}
Let the spectrum of $V$ be simple and let $\{v_{1},\ldots,v_{N}\}$
be the eigenvectors, which form the basis of the space $\mathbb{R}^{N}$.
Then 
\[
\dim L_{0}=2\#\{k\in\{1,\ldots,N\}:\ v_{k}\perp e_{n}\},
\]
where $\perp$ is in the sense of $(,)_{1}$. \end{lemma} \textbf{Proof.}\ As
the subspace $l_{V}$ is invariant with respect to $V$, one can enumerate
the vectors $\{v_{1},\ldots,v_{N}\}$ so that for some $m$ 
\[
l_{V}=\langle v_{1},\ldots,v_{m}\rangle.
\]
Thus, $e_{n}\in\langle v_{1},\ldots,v_{m}\rangle=\langle v_{m+1},\ldots,v_{N}\rangle^{\perp}$,
$\perp$ in the sense of $(,)_{1}$. \ \ \ $\square$

We will calculate the spectrum of the matrix $V$, which acts on the
vectors $q$ as follows: 
\[
(Vq)_{i}=\begin{cases}
\omega_{0}q_{i}-\omega_{1}(q_{i-1}+q_{i+1}-2q_{i}), & 1<i<N-1,\\
\omega_{0}q_{1}-\omega_{1}(q_{2}-q_{1}), & i=1,\\
\omega_{0}q_{N}-\omega_{1}(q_{N-1}-q_{N}), & i=N.
\end{cases}
\]

\begin{lemma} \label{l4} The matrix $V$ has $N$ different eigenvalues
$\lambda_{0},\ldots,\lambda_{N-1}$ , given by 
\[
\lambda_{k}=\omega_{0}+2\omega_{1}\Bigl(1-\cos\frac{\pi k}{N}\Bigr),\quad k=0,\ldots,N-1.
\]
The corresponding eigenvectors $y^{(0)},\ldots,y^{(N-1)}$ have the
following coordinates: 
\[
y_{j}^{(k)}=\cos\pi\frac{(j-1/2)k}{N},\quad j=1,\ldots,N,\ k=0,\ldots,N-1.
\]
\end{lemma}

Lemma~\ref{l4} can be proved by direct substitution.

Using Lemmas~\ref{l3} and \ref{l4} we get 
\begin{align*}
\dim L_{0} & =2\#\{k\in\{0,\ldots,N-1\}:(y^{(k)},e_{n})=0\}\\
 & =2\#\Bigl\{ k\in\{0,\ldots,N-1\}:\ \cos\Bigl(\frac{\pi k(n-1/2)}{N}\Bigr)=0\Bigr\}=\\
 & =2\#\{k\in\{0,\ldots,N-1\}:\ a_{n}(k)\in\mathbb{Z}\},
\end{align*}
where 
\[
a_{n}(k)=\frac{1}{2}\Bigl(\frac{2n-1}{N}k-1\Bigr).
\]
Denote $d=\gcd(N,2n-1)$. For some integers $m_{1},m_{2}$ such that
$\gcd(m_{1},m_{2})=1$, we have 
\[
2n-1=dm_{1},\quad N=dm_{2},\quad a_{n}(k)=\frac{1}{2}\Bigl(\frac{m_{1}}{m_{2}}k-1\Bigr).
\]
As the numbers $d$ and $m_{1}$ are odd, the number $a_{n}(k)$ is
an integer if and only if $k=(2l-1)m_{2},\ l=1,2,\ldots$ From the
condition $k\leqslant N-1$ it follows that 
\[
l\leqslant\frac{1}{2}\Bigl(\frac{N-1}{m_{2}}+1\Bigr).
\]
Thus 
\[
\dim L_{0}=2\Bigl[\frac{1}{2}(\frac{N-1}{N}d+1)\Bigr]=2\Bigl[\frac{d}{2}-\frac{d}{2N}\Bigr].
\]
Since $(d/2N)\leqslant(1/2)$ and $d$ is odd, we have 
\[
\Bigl[\frac{d}{2}-\frac{d}{2N}\Bigr]=\frac{d-1}{2}.
\]
This gives the first statement of Theorem~\ref{t2}. Statements 2
and 3 of the theorem follow from the results of \cite{Brough}, see
theorems 2.1, 3.2, 4.2 therein.

\end{document}